\documentclass[hyper]{prop2015}% no class options needed by now
\usepackage[english]{babel}
\usepackage{bbm}
\usepackage{amsmath}
\DeclareMathOperator{\tr}{tr}

\category{Proceedings}
\subtitle{\href{http://www.maths.dur.ac.uk/lms/109/index.html}{LMS/EPSRC Durham Symposium on Higher Structures in M theory}}
\title{M-Branes: Lessons from M2's and Hopes for M5's}
\author[N. Lambert]{Neil Lambert\inst{a,}\footnote{Corresponding author e-mail:~\href{mailto:neil.lambert@kcl.ac.uk}{\textsf{neil.lambert@kcl.ac.uk}}}}
 \address[1]{Department of Mathematics, King's College London, The Strand, WC2R 2LS, United Kingdom}
\shortauthors{N. Lambert}
\begin{abstract}
In this talk we will review the construction of M2-brane SCFT's highlighting some novelties and the role of 3-algebras. Parts of our discussion will closely follow parts of \cite{Bagger:2012jb}. Next we will discuss M5-branes: the basics, the obstacles as well as various attempts to construct the associated SCFT and potential relations between M2-branes and M5-branes. 
\end{abstract}
\shortabstract
\begin{document}
\maketitle
%%% Use this if the article text won't start with a \section:
% \noindent
%%% Being based on LaTeX's article class, and2012 supports the respective 
%%% sectioning level from \section to \subparagraph.

\section{Introduction}
  
I was asked to give a review talk on the construction of the M2-brane SCFT's and also to detail the issues and problems associated to the infamous M5-brane SCFT's. Not wanting to feel left out I also tried to add something of my own recent work which I hoped would be of interest to my colleagues at the conference. Therefore 
the plan of this talk is split into three themes:
\bigskip
\begin{enumerate}[i)]
\item M2-branes and 3-algebras
\item M5-branes and the $(2,0)$ theory
 \item A $(2,0)$ system
\end{enumerate}
 \bigskip
The aim of the first theme is to review the construction of the M2-brane Chern--Simons SCFT's with a view to emphasising the role of 3-algebras. However I also want to point out that although we have Lagrangian descriptions for arbitrary numbers of M2-branes in many eleven-dimensional backgrounds,  these Lagrangians do not have all the symmetries that one expects. Instead these only arise in the quantum theory at strong coupling through non-perturbative operators. In the second theme I will discuss the objections to obtaining a Lagrangian description of M5-branes and review a few attempts to define the associated SCFT using lower dimensional Lagrangian theories. Finally in the remaining theme I would like to present an explicit representation of the $(2,0)$ super-algebra on a set of fields and show that by making certain choices for the solutions to the constraints one recovers various Lagrangian descriptions of M5-branes and M2-branes.

On the other hand there is much work on M5-branes that I will not discuss. Not out of a lack of interest but out of a lack of time and knowledge. Amongst the plethora of work that I will not mention are:
\bigskip  
\begin{enumerate}[i)]
 \item Results arising from reduction to 4D and below such as novel non-Lagrangian  field theories, dualities, surface operators, AGT  etc
(e.g. \cite{Gaiotto:2009we}, $\ldots$)
 \item  Bootstrap results for M5-branes (e.g. \cite{Beem:2015aoa}, $\ldots$)
\item   AdS$_7$/CFT$_6$   (e.g. \cite{Heslop:2017sco,Chester:2018dga}, $\ldots$)
 \end{enumerate}
 \bigskip  
 The moral of this talk is that although M2-branes are essentially a done deal there are details in the fine print that could hold lessons for M5-branes. And furthermore although an explicit M5-brane construction from some Lagrangian-like system seems unlikely there is still hope that novel techniques and physics can emerge and that we will learn new things about quantum field theory.

  \section{M2-branes}

The M2-brane SCFT arises as the strong coupling limit of $N$ D2-branes. These are described in the decoupling limit by 3D maximally supersymmetric  Yang--Mills (MSYM)  with gauge group $U(N)$. The strong coupling limit corresponds to the IR limit. However the lift to M-theory implies that at strong coupling an extra eleventh dimension arises and the R-symmetry is thereby increased from $SO(7)$ to $SO(8)$:
\begin{equation}
SO(1,2)_L\times SO(7)_R \to SO(1,2)_L\times SO(8)_R\ ,
\end{equation}
here $L$ stands for Lorentz symmetry and R for R-symmetry. Ultimately these are subgroups of the ten and eleven-dimensional Lorentz groups.
Therefore the standard type IIA/M-theory dictionary predicts that there is a   3D SCFT with maximal supersymmetry and $SO(8)$ R-symmetry corresponding to the IR limit of 3D super-Yang--Mills. Thus although we started with string theory and M-theory we have reached a conclusion that is simply about gauge theory and QFT. A prediction so to speak.

The relevant Lagrangians for these  theories have now been constructed. The first example with maximal (${\cal N}=8$)  manifest supersymmetry is BLG \cite{Bagger:2007jr,Gustavsson:2007vu}. It is a Chern--Simons-matter theory with gauge group  $SU(2)\times SU(2)$ or $(SU(2)\times SU(2))/{\mathbbm Z}_2$. However it is limited in that it only describes two or three M2's on an orbifold.

 For arbitrary number of M2-branes one has ${\cal N}=6$ manifest  SUSY and the ABJM  or ABJ  models \cite{Aharony:2008ug,Aharony:2008gk}. Here one gives up manifest maximal supersymmetry and instead has only 12 supercharges. It is again a Chern--Simons-matter theory but with gauge group  $ U(M)\times U(N)$ and it describes  $N\le M$ branes in an eleven-dimensional orbifold. 
 
 There is now a zoology of Chern--Simons-matter theories with extended SUSY ${\cal N}=4,5,6,8$ corresponding to  a motley list of gauge groups. For ${\cal N}=3$ there is no restriction on the gauge group \cite{Kao:1992ig}.
 
\subsection{3-algebras} 
  
A central ingredient to all these theories is a 3-algebra. This is a vector space ${\cal V}$ with a triple product 
\begin{equation}
[\ \cdot\ ,\ \cdot\  ,\ \cdot\  ]:{\cal V}\otimes {\cal V}\otimes {\cal V} \to {\cal V}\ ,
\end{equation}
such that the endomorphism $\varphi(\ \cdot \ )= [\ \cdot\ ,U ,V]:{\cal V}  \to {\cal V}$, for fixed $U,V\in {\cal V}$, is a derivation. This leads to the so-called fundamental identity:
\begin{equation}
\varphi([A,B,C]) = [\varphi(A),B,C]+ [A,\varphi(B),C]+ [A,B,\varphi(C) ]\ .
\end{equation}
For physics we require that there is a positive definite inner product on ${\cal V}$:
\begin{equation}
\langle \ \cdot\ ,\ \cdot \ \rangle: {\cal V}\otimes {\cal V} \to {\mathbbm  R}  \ ,
\end{equation}
which induces an invariant inner product on the space of derivations:
\begin{equation}
(T,\varphi) = \langle T(U),V \rangle \ .
\end{equation}

There is also a complex version  of  a 3-algebra:
\begin{equation}
[\cdot,\cdot;\cdot]:{\cal V}\otimes {\cal V}\otimes \bar{\cal V} \to {\cal V}\ ,
\end{equation}
with complex positive definite inner product:
For physics we require that there is a positive definite inner product on ${\cal V}$:
\begin{equation}
\langle \ \cdot\ ,\ \cdot \ \rangle: {\cal V}\otimes {\cal V} \to {\mathbbm  C}\ .  
\end{equation}
Again the analogue of adjoint map
\begin{equation}
\varphi_{U,\bar V}(X)=[X,U;\bar V]~,\qquad  \varphi_{U,\bar V}(\bar X)=-[\bar X,\bar V; U]\ ,
\end{equation}
is a derivation
\begin{align}
 &\varphi_{U,\bar V}([X,Y;\bar Z]) =\\
 &\kern.3cm= [\varphi_{U,\bar V}(X),Y;\bar Z] + [X,\varphi_{U,\bar V}(Y);\bar Z]\nonumber + [X,Y;  { \varphi_{U,\bar V}(\bar Z)}]\ .
\end{align}
 
The fundamental identity tells us that the action of $\varphi$ on $\cal V$ is that of a Lie algebra $\mathfrak{g}$ generated by $\varphi_{U, V}$ for all $U,V\in {\cal V}$. In other words ${\cal V}$ is representation of $\mathfrak{g}$. Thus
  a 3-algebra defines a Lie algebra  $\mathfrak{g}$ along with a preferred representation.

In fact the reverse is also true: given a Lie algebra and a representation (along with invariant inner products) one can always construct a triple product satisfying the fundamental identity via the so-called {Faulkner} map. Such 3-algebras, including ones with mixed signature inner products (which also have applications to gauge theory) have been classified, see for example \cite{deMedeiros:2008zh,deMedeiros:2009hf}

One need not think of a  3-algebra   and just think of the gauge group and matter representation. However the triple product fixes all the terms in the Lagrangian.  Furthermore the amount of manifest supersymmetry fixes the symmetry properties of the triple product which in turn restricts the choice of 3-algebra and  hence which gauge algebras and representations arise. This is a rather novel situation as  the amount of manifest supersymmetry is determined by the gauge algebra and matter representations, unlike the case of 
  super-Yang--Mills theories where the gauge group is arbitrary. Furthermore even though the gauge fields are related to the matter fields by supersymmetry    they do not sit in the same representation of the gauge group. This is possible as the Chern--Simons structure means that the gauge fields do not carry on-shell degrees of freedom. 
  
 \subsection{Examples} 
  
  Let us look at some examples.  
  
 \subsubsection{${\cal N}=8$ supersymmetry: BLG}
 
 We take ${\cal V}$ real  and $[\ \cdot\ ,\ \cdot\ ,\ \cdot\ ]$ totally antisymmetric and
 \begin{subequations}
 \begin{eqnarray}
\delta X^I  &=& i\bar\epsilon\Gamma^I\Psi \\
\delta \Psi  &=& D_\mu X^I \Gamma^\mu \Gamma^I\epsilon -\frac{1}{6}
[X^I ,X^J, X^K] \Gamma^{IJK}\epsilon \\
 \delta  A_{\mu}(\ \cdot\ ) &=& i\bar\epsilon
\Gamma_\mu\Gamma_I[\ \cdot \ ,X^I,\Psi ]\; , 
\end{eqnarray}
\end{subequations}
(here $\mu,\nu=0,1,2$ and $I,J=3,4,5,\ldots,10$).
The Lagrangian is 
\begin{equation}
\begin{aligned}
{\cal L}  = &-\frac{1}{2}\langle D_\mu X^{ I}D^\mu X^{I} \rangle+\frac{i}{2}\langle\bar\Psi, \Gamma^\mu D_\mu \Psi\rangle\,+\\ 
& +\frac{i}{4}\langle\bar\Psi, \Gamma_{IJ}[X^I,X^J,\Psi ]\rangle\,- \\
 &- \frac{1}{12}\langle[X^I,X^J,X^K ],[X^I,X^J,X^K] \rangle\,+\\ 
 & +\varepsilon^{\mu\nu\lambda}\left( (A_{\mu}, \partial_\nu A_{\lambda  }) +\frac{1}{3} (A_{\mu  },[A_{\nu },A_{\lambda  }])\right) \ .
 \end{aligned}
 \end{equation}

But for a positive definite choice of inner product (which we take to be the identity) there is just one finite-dimensional solution \cite{Gauntlett:2008uf,Papadopoulos:2008sk}:
\begin{equation}
[T^a,T^b,T^c] = \frac{4\pi}{k}\varepsilon^{abcd}T^d\qquad a,b,c,d=1,2,3,4\ .
\end{equation}
The gauge algebra generated by $\varphi$ is $\mathfrak{so}(4)=\mathfrak{su}(2)_+\oplus \mathfrak{su}(2)_-$ and
\begin{equation}
((T^+,T^-),(W^+,W^-)) = \frac{k}{4\pi}{\rm tr}(T^+W^+)-\frac{k}{4\pi}{\rm tr}(T^-W^-)\ ,
\end{equation}
so we find an $\mathfrak{su}(2)_+\oplus \mathfrak{su}(2)_-$ Chern--Simons Lagrangian with opposite levels: 
\begin{equation}
\begin{aligned}
{\cal L}_{CS} &=  \frac{k}{4\pi}\varepsilon^{\mu\nu\lambda}{\rm tr}\left(  A^+_{\mu
}  \partial_\nu A^+_{\lambda  }  +\frac{1}{3}  
A^+_{\mu  }[A^+_{\nu },A^+_{\lambda  }]\right)- \\
  &\kern.5cm - \frac{k}{4\pi}\varepsilon^{\mu\nu\lambda}{\rm tr} \left(A^-_{\mu
}  \partial_\nu A^-_{\lambda  }) +\frac{1}{3} (
A^-_{\mu  }[A^-_{\nu },A^-_{\lambda  }])\right)\ .
\end{aligned}
\end{equation}
The fields $X^I$, $\Psi$ are in the $\bf 4 = \bf 2 +\overline  {\bf 2}$ = bifundamental. A standard result tells us that $k\in \mathbbm Z$ - no continuous parameter. 

\subsubsection{${\cal N}=6$ supersymmetry: { ABJM}}

 We need a little less symmetry and a complex $\cal V$.  To this end we write $X^I $   as four complex scalar fields $Z^A $ $A=1,2,3,4$ in $\bf 4$ of $SU(4)$ with $U(1)$ charge 1. And $\Psi $ is now written as 4 complex fermions $\psi_{A}$ in   $\overline  {\bf 4}$ with $U(1)$ charge 1. Lastly the
16 components of $\epsilon$ are reduced to $\epsilon^{AB} =- \epsilon^{BA}$ in $\bf 6$ of $SU(4)$ with $U(1)$ charge 0. We can now write down the supersymmetry transformations:
\begin{subequations}
\begin{eqnarray}
  \delta Z^A  &=& i\bar\epsilon^{AB}\psi_{B }\\
  \delta \psi_{B } &=& \gamma^\mu D_\mu Z^A \epsilon_{AB} +
 [Z^C, Z^A; {\bar Z}_{C}] \epsilon_{AB}\,+\nonumber\\
&&\kern.5cm+\,[Z^C, Z^D; {\bar Z}_{B}]\epsilon_{CD} \\
  \delta   A_\mu{}  (\ \cdot \ )&=&
i\bar\epsilon_{AB}\gamma_\mu[  \ \cdot\ , Z^A ;\psi^{B}]  -
i\bar\epsilon^{AB}\gamma_\mu [ \ \cdot\ {\bar Z}_{A}  ; \psi_{B }] \ , 
\end{eqnarray}
\end{subequations}
and Lagrangian
\begin{subequations}
\begin{equation}\label{nsixlag}
\begin{aligned}
 {\cal L} &= - \langle D^\mu {\bar Z}_A , D_\mu Z^A \rangle - i\langle\bar\psi^{A }\gamma^\mu , D_\mu\psi_{A }\rangle -V \,-\\
&\kern.5cm -\,i  \langle\bar\psi^{A }, [\psi_{A } , Z^B;{\bar Z}_{B}]\rangle+ 2i \langle\bar\psi^{A },[\psi_{B } ,Z^B; {\bar Z}_{A}]\rangle\,+\\
&\kern.5cm+\,\frac{i}{2}\varepsilon_{ABCD}\langle \bar\psi^{A},[Z^C;Z^D;\psi^{B}]\rangle\,-\\
&\kern.5cm-\,\frac{i}{2}\varepsilon^{ABCD} \langle\bar\psi_{A},[{\bar Z}_{C},{\bar Z}_{D};\psi_{B} ]\rangle\,+\\
&\kern.5cm +\,\varepsilon^{\mu\nu\lambda}\left( A_{\mu}  ,\partial_\nu A_{\lambda} \right)+\frac{1}{3}\varepsilon^{\mu\nu\lambda}\left(A_{\mu} , [ A_{\nu} , A_{\lambda}]\right)\ ,
\end{aligned}
\end{equation}
where the potential is  
\begin{equation}
V = \frac{2}{3}\langle\Upsilon^{CD}_{B},\bar\Upsilon_{CD}^{B } \rangle
\end{equation}
with
\begin{equation}
\Upsilon^{CD}_{B} = [Z^C,Z^D,{\bar Z}_{B}]-\frac{1}{2}\delta^C_B[Z^E,Z^D;{\bar Z}_{E}]+\frac{1}{2}\delta^D_B[ Z^E,Z^C;{\bar Z}_{E}] 
\end{equation}
\end{subequations}
An infinite class of solutions are given by $M\times N$ complex matrices with $\langle A,B\rangle = {\rm tr}(A B^\dag)$ and 
\begin{equation}
[A,B;C] = \frac{4\pi}{k}(AC^\dag B - B C^\dag A)\ .
\end{equation}
The gauge transformation generated by $\delta Z^A  =[Z^A,U,\bar V]$ is
\begin{equation}
\delta Z^A = MZ^A - Z^AN\ ,
\end{equation}
where $M=-V^\dag U ,N = UV^\dag $ are $M\times M$ and $N\times  N$ matrices respectively and
\begin{equation}
(M,M') = \frac{k}{4\pi}{\rm tr}(MM')\qquad (N,N') = -\frac{k}{4\pi}{\rm tr}(NN')\ .
\end{equation}
Thus we find the gauge group   $U(M)\times U(N)$ with matter in the bi-fundamental. Cases with $M > N$ are known as the {ABJ} theories. For $N=M$ one actually just finds $SU(N)\times SU(N)$ but the missing $U(1)$ factors can be added in by hand as they are supersymmetry singlets.
In the special case of $SU(2)\times SU(2)$ we recover the {BLG} theory in complex notation.

The list of examples continues with less supersymmetry depending on the symmetry properties of the structure constants
\begin{equation}
[T^a,T^b;T_c] = f^{ab}{}_{cd}T^d\ ,
\end{equation}
but the actions are essentially the same.

\subsection{Novelties } 
 As we have mentioned above these actions `break'  some supersymmetry `rules'.
 \bigskip  
 \begin{enumerate}[i)]
 \item Gauge fields and matter fields are in the same multiplet but not in the same representation of the gauge group. 
 \item The amount of supersymmetry is determined by the gauge group
 \end{enumerate}
 \bigskip  
In particular (for example see \cite{Bagger:2010zq}):
\begin{figure}
\includegraphics[width=\columnwidth]{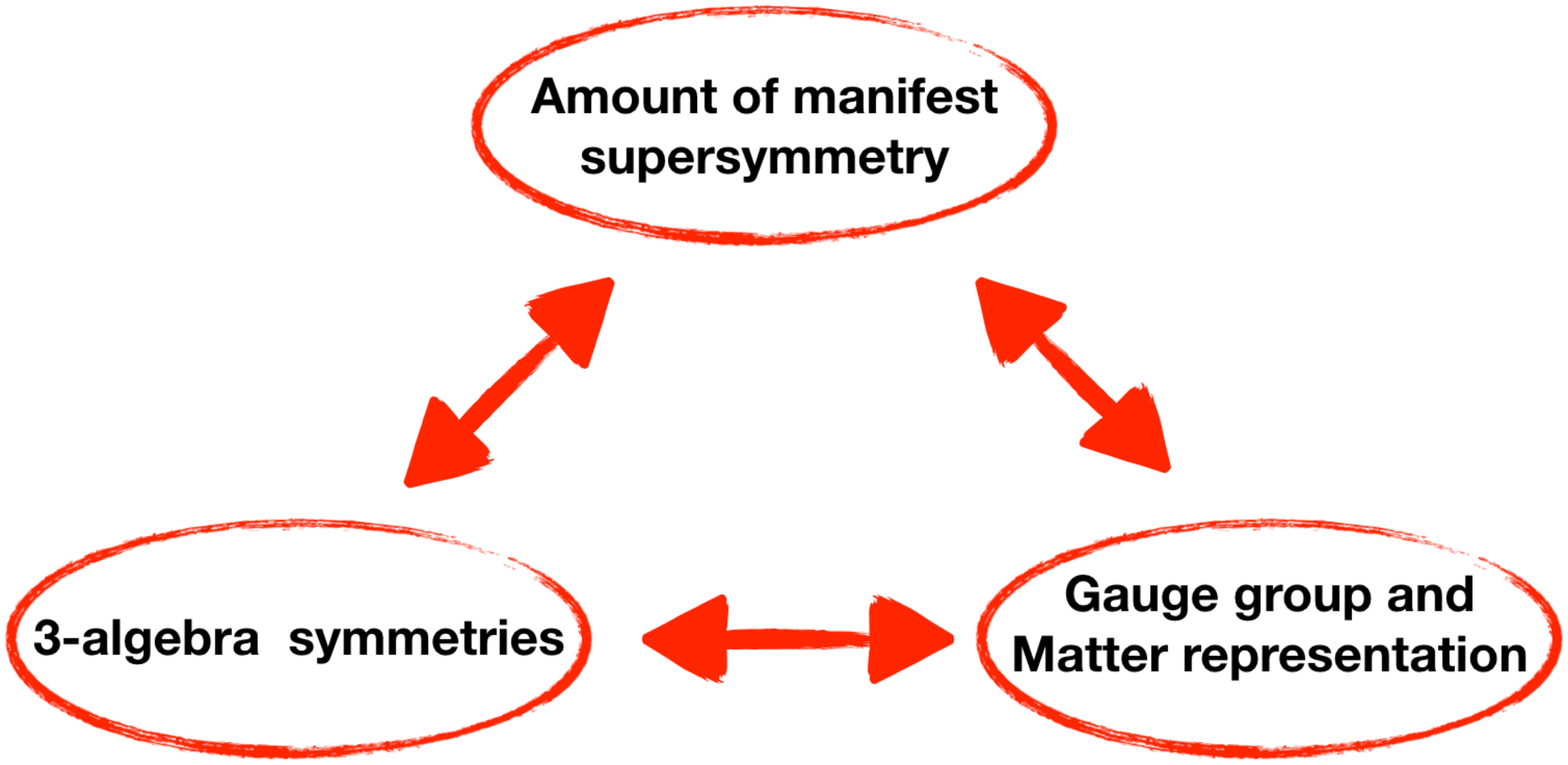}
\caption{\label{3repsusy}{SUSY, 3-algebras and representations}}
\end{figure} 
\begin{subequations}
  \begin{align}
 f^{abcd} = f^{[abcd]}~\Longleftrightarrow~ &\mathcal{N}=8 ~\Longleftrightarrow~ \mathfrak{su}(2)\oplus \mathfrak{su}(2)\\
 \left(\begin{array}{c} 
 f^{ab}{}_{cd} = f^{[ab]}{}_{cd}\\   f^{ab}{}_{cd}=(f^{cd}{}_{ab})^* 
 \end{array}\right)
~\Longleftrightarrow~ &\mathcal{N}=6 ~\Longleftrightarrow~    \begin{array}{c} \mathfrak{u}(N)\oplus \mathfrak{u}(M)   \\ \mathfrak{sp}(N)\oplus \mathfrak{u}(1) \end{array} \\\
  \left(\begin{array}{c} 
 f^{abcd}  = f^{[ab] cd}\\   f^{abcd}=(f_{abcd})^*  
 \end{array}\right)
~\Longleftrightarrow~ &\mathcal{N}=5 ~\Longleftrightarrow~ 
 \begin{array}{c} \mathfrak{sp}(N)\oplus \mathfrak{su}(M)   \\ \mathfrak{so}(7)\oplus \mathfrak{su}(2)\\ \quad  \mathfrak{g}_2\oplus \mathfrak{su}(2)\end{array} 
 \end{align}
 \end{subequations}
 The 3-algebra formalism is a neat way of encoding all this data even though in the end one is always just talking about a Chern--Simons-matter field theory based on a gauge group and choice of representation.

  \subsection{Essential dynamics} 

Having constructed these theories it begs the question as to whether or not they really describe M2-branes. For a start there is no obvious free centre of mass multiplet. I warn you now that his is a rather lengthy and detailed section so please bare with me (or skip to the end). I am mentioning it here to help illustrate some points later: namely that one has to work hard, and within the quantum theory, to see the correct physics.

The first thing to look at is the vacuum moduli space. This tells us the space of all the zero-energy configurations of the M2-branes. We will just stick to
ABJM:
\begin{equation}
[Z^A,Z^B;\bar Z_C]=0 ~\Longleftrightarrow~ Z^A\bar Z_C Z^B = Z^B\bar Z_C Z^A\ .
\end{equation}
Generically this implies that all the $Z^A$ commute ({\it c.f.} D-branes):
\begin{equation}
Z^A = {\rm diag}(z_1^A,\ldots,z_n^A)\ .
\end{equation}
To see that this is all requires one to evaluate the mass formula for small fluctuations which one finds is non-zero (generically: there are special points where extra massless modes arise but are expected to be lifted by non-perturbative effects).

We must identify fields that differ by gauge transformations:
\begin{equation}
Z^A \to g_L Z^A g_R^{-1}\ .
\end{equation}
We could set $g_L=g_R$ so that this is an adjoint action, as with D-branes. This allows us to put $Z^A$ in diagonal form (as we have already done) and in addition acts as
\begin{equation}
z^A_i \leftrightarrow z^A_j\qquad {\rm for\ any \ } i\ne j\ ,
\end{equation}
e.g. for $i,j=1,2$ these are generated by
\begin{equation}
g_L=g_R = \left(
\begin{array}{ccccc}
 0& i  & & & \\
 i & 0 &    &&\\
  &&1&&\\
  & &   &\ddots   &\\
  & & &&1
 \end{array}   \right)\ .
\end{equation}
These generate the action of the symmetric group $S_N$ on $z^A_i$.

Unlike D-branes we also have continuous gauge transformations:
\begin{equation}
z^A_i \to e^{i\theta_i}z^A_i\ .
\end{equation}
These arise from taking
\begin{equation}
g_L= g_R^{-1}= {\rm diag}(
 e^{i\theta_1/2},\ldots,e^{i\theta_N/2} )\ .
\end{equation}
To see the effect of this on the vacuum moduli space we must examine the Lagrangian for the moduli $z^A_i$, including the gauge fields.  The Lagrangian on the moduli space is
\begin{subequations} 
\begin{equation}
\begin{aligned}
{\cal L} = &-\frac{1}{2}\sum_i D_\mu z^A_i D^\mu \bar z_{Ai}\,+\\ 
&\kern.5cm+\, \frac{k}{4\pi}\varepsilon^{\mu\nu\lambda}\sum_iA^L_{ \mu i}\partial_\nu A^L_{ \lambda i}- \frac{k}{4\pi}\varepsilon^{\mu\nu\lambda}\sum_iA^R_{ \mu i}\partial_\nu A^R_{ \lambda i}\ ,
\end{aligned}
\end{equation}
where
\begin{equation} 
A^L_\mu = {\rm diag}(A_{\mu 1}^L,\ldots,A_{\mu N}^L)$, $  A^R = {\rm diag}(A_{\mu 1}^R,\ldots,A_{\mu N}^R)\ ,
\end{equation}
and
\begin{equation}
D_\mu z^A_i = \partial_\mu z^A_i - i(A^L_{\mu i}-A^R_{\mu i})z^A_i\ .
\end{equation}
\end{subequations}
 Note that $z^A_i$ only couples to $B_{\mu i } = A^L_{\mu i } - A^R_{\mu i }$ and not to $Q_{\mu i}= A^L_{\mu i } + A^R_{\mu i }$:
\begin{equation}
{\cal L} = -\frac{1}{2}\sum_i D_\mu z^A_i D^\mu \bar z_{Ai} + \frac{k}{4\pi}\varepsilon^{\mu\nu\lambda}\sum_iB _{ \mu i}\partial_\nu Q _{ \lambda i}\ ,
\end{equation}
with $D_\mu z^A_i = \partial_\mu z^A_i - iB_{\mu i} z^A_i$.

It is helpful to dualize $Q_{\mu i}$ by introducing a Lagrange multiplier $\sigma_i$:
\begin{equation}
\begin{aligned}
{\cal L} 
%&=& -\frac{1}{2}\sum_i D_\mu z^A_i D^\mu \bar z_{Ai}\nonumber\\ &&+ \frac{k}{8\pi}\varepsilon^{\mu\nu\lambda}\sum_iB _{ \mu i}H_{ \nu\lambda i} - \frac{1}{8\pi}\varepsilon^{\mu\nu\lambda}  \sigma_i \partial_\mu H_{\nu\lambda i}\nonumber\\
&\cong -\frac{1}{2}\sum_i D_\mu z^A_i D^\mu \bar z_{Ai}\,+ \\ 
&\kern.5cm+ \frac{k}{8\pi}\varepsilon^{\mu\nu\lambda}\sum_iB _{ \mu i}H_{ \nu\lambda i} + \frac{1}{8\pi}\varepsilon^{\mu\nu\lambda} \partial_\mu \sigma_i H_{\nu\lambda i}\ ,
\end{aligned}
\end{equation}
where $H_{ \nu\lambda i}  = \partial_\nu Q_{ \lambda i} -\partial_\lambda Q_{ \nu i}  $.

Integrating out $H_{\nu\lambda i}$ tells us $B_{\mu i} = -k^{-1}\partial_{\mu}\sigma_i$ and everything is pure gauge:
\begin{equation}
{\cal L} = -\frac{1}{2}\sum_i \partial_\mu  w^A_i  \partial^\mu  \bar w_{Ai}\ ,  
\end{equation}
where $w^A_i = e^{i\sigma_i/k}z^A_i$ is gauge invariant.
 Next we observe that  $\sigma_i$ is periodic:
\begin{eqnarray}
\int {\cal L}(\sigma_i +2\pi) -\int {\cal L}(\sigma_i) &=& -\frac{1}{4}\sum_i\int \varepsilon^{\mu\nu\lambda}   \partial_\mu H_{\nu\lambda i}\nonumber\\
 &=&  - \frac{1}{2}\sum_i\int dH \nonumber \\
 &=&   - \frac{1}{2}\sum_i\int dF^L+ dF^R\nonumber\\
 &\in & 2\pi {\mathbbm Z}\ ,
 \end{eqnarray}
because of the Dirac quantization rule
\begin{equation}
\int {\rm d}F  \in  2\pi {\mathbbm Z}\ ,
\end{equation}
as well as the fact that $B_i = -k^{-1}{\rm d}\sigma_i$ implies ${\rm d}B_i = F^L_i-F^R_i=0$.\footnote{NB: This is very sensitive to the global choice of gauge group $\mathfrak{u}(N),~\mathfrak{su}(N),~\mathfrak{su}(N)/{\mathbbm Z}_N$.}
This means that (recall $w^A_i = e^{i\sigma_i/k}z^A_i$)
\begin{equation}
w^A_i \cong e^{2\pi i/k}w^A_i\ .
\end{equation}
Thus there is an extra orbifold action in space-time
\begin{equation}
{\mathbbm R}^8 \to {\mathbbm C}^4/{\mathbbm Z}_k\ ,
\end{equation}
and the vacuum moduli space is
\begin{equation}
{\cal M}  = {\rm Sym}^n\left({\mathbbm C}^4/{\mathbbm Z}_k\right)\ .
\end{equation}
Corresponding to $N$ M2-branes in an ${\mathbbm C}^4/{\mathbbm Z}_k$ transverse space. And indeed and M2-brane in this orbifold preserves 12 supersymmetries.
This explains why there is no translational mode for generic $k$ (including the classical, large $k$, limit). But it should be there for $k=1$ and we will find it later.
 
Let us return to the moduli space. It follows that we can think of
\begin{equation}
Z^A = \left(
\begin{array}{ccc }
 z_i^A&   &  \\
  &  \ddots   &\\
  & &z^A_n
 \end{array}
  \right)\ ,
\end{equation}
as describing the positions of $N$ M2-branes in ${\mathbbm C}^4/{\mathbbm Z}_k$. Furthermore the natural circle for the M-theory direction is the over-all phase.

Suppose we wanted to describe $N$ M2-branes moving along the M-theory circle with different speeds. One might expect that this corresponds to
\begin{equation}
Z^A = \left(
\begin{array}{ccc }
 z_i^Ae^{i\omega_1 t}&    &  \\
  &  \ddots   &\\
  & &z^A_N e^{i\omega_N t}
 \end{array}
  \right)\ .
\end{equation}
But this is pure gauge! We can un-do it by taking\footnote{NB: This gauge transformation is not allowed for D-branes where the scalars are in the adjoint.}
\begin{equation}
g_L= g_R^{-1}=\left(
\begin{array}{ccc }
  e^{-i\omega_1 t/2}&    &  \\
  &  \ddots   &\\
  & & e^{-i\omega_N t/2}
 \end{array} 
  \right)\ .
\end{equation}
So how do the M2-branes `explore' the full transverse space? Let us set the fermions to zero and construct the Hamiltonian
\begin{equation}
\begin{aligned}
H=  &\tr \int {\rm d}^2 x\   \Pi_{Z^A}\Pi_{\bar Z_A} + D_i Z^A D^i\bar Z_A  +V\,+\\
&\kern.5cm+\,   \left(iZ^A\Pi_{Z^A}-i\Pi_{\bar Z_A}\bar Z_A  - \frac{k}{2\pi}F^L_{12}\right)A^L_0\,+\\
&\kern.5cm+\,    \left(i\bar Z_A\Pi_{\bar Z_A} -i\Pi_{Z^A}Z^A  + \frac{k}{2\pi}F^R_{12}\right)A^R_0\ .
\end{aligned}
\end{equation}
As usual the time-components of the gauge field give constraints:
\begin{equation}
\begin{aligned}
\frac{k}{2\pi}F^L_{12} &= iZ^A\Pi_{Z^A}-i\Pi_{\bar Z_A}\bar Z_A~,  \\
  \frac{k}{2\pi}F^R_{12}&=i\Pi_{Z^A}Z^A-i\bar Z_A\Pi_{\bar Z_A} \ .
  \end{aligned}
  \end{equation}
Let us consider the vacuum moduli again:
\begin{equation}
Z^A = \left(
\begin{array}{ccc }
  \frac{1}{\sqrt{2}}R^A_1e^{i\theta_1^A}&    &  \\
  &  \ddots   &\\
  & &\frac{1}{\sqrt{2}}R^A_ne^{i\theta_n^A}
 \end{array}
  \right)\ .
\end{equation}
The constraint is
\begin{equation}
\frac{k}{2\pi}F^L_{12} = \frac{k}{2\pi}F^R_{12} = \left(
\begin{array}{ccc }
  \sum_A(R^A_1)^2\partial_0\theta^A_0&    &  \\
  &  \ddots   &\\
  & &\sum_A(R^A_n)^2\partial_0\theta^A_n
 \end{array}   \right)\ .
\end{equation}
In other words the momentum around the M-theory circle is given by the magnetic flux. In spirit this is the same as dualization:
\begin{equation}
\partial_\mu X^{10} = \frac{1}{2}\varepsilon_{\mu\nu\lambda}F^{\nu\lambda}\qquad \Longrightarrow \qquad \partial_0 X^{10} = F_{12}\ .
\end{equation}

This raises the next question: how do we compute quantities with eleven-dimensional momentum. In particular the gauge invariant observables appear to only carry vanishing $U(1)$ charges:
\begin{equation}
\begin{aligned}
{\cal O} &=  \tr(Z^A\bar Z_BZ^C\cdots)\qquad {\rm \color{red}{OK}}~,\\
{\cal O} &=  \tr(Z^A Z^BZ^C\cdots)\qquad {\rm \color{red}{not~OK}}
\end{aligned}
\end{equation}
and hence don't really explore all eleven dimensions.

This brings us to monopole or 't Hooft operators: we want to create states that carry magnetic charge. These operators are defined as a prescription for computing correlators in the path integral. They are not constructed as a local expression of the fields. In particular, a monopole operator ${\cal M}(y)$ is defined by modifying the boundary conditions of the fields about the point $y$ in the path integral
\begin{equation}
\langle{\cal M}(y){\cal O}(z)\cdots\rangle = \int_{\frac{1}{2\pi}\oint_y F =   Q_M} DZD\psi DA {\cal O}(z)e^{-S}\ ,
\end{equation}
in other words we require the fields in the path integral to have a specific  singularity
\begin{equation}
F = \star \frac{Q_m}{2}{\rm d}\left(\frac{1}{|x-y|}\right)  + {\rm nonsingular}\ .
\end{equation}
$Q_M\in\mathfrak{u}(n)\times\mathfrak{u}(n)$ is the magnetic flux and is subject to the standard Dirac quantization condition
\begin{equation}
e^{2\pi i Q_m}  = 1\ .
\end{equation}
 
Next we note that due to the Chern--Simons term   monopole operators transform locally under a gauge transformation $\delta A^{L/R}_\mu  =D_\mu\omega_{L/R}$ (with $\omega\to 0$ at infinity) as
 \begin{eqnarray}
 {\cal M}_{Q_M}(x)  & \to & e^{(ik/2\pi)\, {\rm tr} \int ( D\omega_L \wedge  F^L-D\omega_R\wedge F^R)}  {\cal M}_{Q_M}(x)\nonumber \\
& = &e^{i k\, {\rm tr}((\omega_L(x)-\omega_R(x)) Q_M)}{\cal M}_{Q_M}(x)\ .
\end{eqnarray}
Note that by construction we have broken the gauge group to $U(1)^n\times U(1)^n$.
This is enough to tell us that under full gauge transformations the monopole operators  transform in the representation of $U(n)\times U(n)$ whose highest weight is
\begin{equation}
\vec\Lambda = k(\vec Q_m,-\vec Q_m)\ ,
\end{equation}
(actually because of the sign the second factor is the lowest weight).
 
This is all very abstract (and tricky to calculate with). Consider the Abelian case (from the moduli space calculation and Wick rotated):
\begin{equation}
\begin{aligned}
{\cal L} = &-\frac{1}{2}\sum_i D_\mu z^A_i D^\mu \bar z_{Ai}\,+\\
 &\kern.5cm+\, \frac{k}{8\pi}\varepsilon^{\mu\nu\lambda}\sum_iB _{ \mu i}H_{ \nu\lambda i} - \frac{i}{8\pi}\varepsilon^{\mu\nu\lambda} \sigma_i \partial_\mu H_{\nu\lambda i}\ .
\end{aligned}
\end{equation}
The monopole operators are just
\begin{equation}
{\cal M}_i(y) = e^{i\sigma_i(y)}\ ,
\end{equation}
since
\begin{equation}
\begin{aligned}
&\langle{\cal M}_i(y){\cal O}(z)\cdots\rangle =\\
&\kern.5cm= \int DzDBDQ e^{i\sigma_i(y)}{\cal O}(z) e^{-\int {\rm d}^3x {\cal L}(x)}\\
 &\kern.5cm= \int DzDBDQ  {\cal O}(z) e^{-\int {\rm d}^3x ({\cal L}(x) + i\sigma_i(x)\delta(x-y))}\ .
 \end{aligned}
\end{equation}
This is the same as taking
\begin{equation}
\frac{1}{8\pi} \varepsilon^{\mu\nu\lambda}   \partial_\mu H_{\nu\lambda i}\to \frac{1}{8\pi} \varepsilon^{\mu\nu\lambda}   \partial_\mu H_{\nu\lambda i} + 8\pi\delta (x-y)\ ,
\end{equation}
i.e. inserting a magnetic charge at $x=y$.

Thus our gauge invariant operator on the moduli space is just
\begin{equation}
w^A_i = e^{i\sigma_i/k}z^A_i = ({\cal M}_i)^\frac{1}{k} z^A_i\ ,
\end{equation}
and indeed ${\cal M}_i$ has charge $(k,-k)$ under $U(1)\times U(1)$.

Even at $k=1$   translations in the transverse space are not symmetries of the Lagrangian:
\begin{equation}
P^A_\mu  = {\rm tr}(D_\mu Z^A)\ ,\qquad {\color{red} {\rm not\ OK}}
\end{equation}
But now we can construct the conserved current (but only for $k=1$):
\begin{equation}
P^A_\mu  = {\rm tr}({\cal M}_{\vec\lambda_1,-\vec\lambda_1}D_\mu Z^A)\ ,\qquad  {\color{red}\rm  OK} 
\end{equation}
as well as the additional two supersymmetries that enhance ${\cal N}=6 \to {\cal N}=8 $:
\begin{equation}
S_\mu  = {\rm tr}({\cal M}_{2\vec\lambda_1,-2\vec\lambda_1}\Psi_A D_\mu Z^A)\ .\qquad {\color{red} {\rm OK}}
\end{equation}

Finally we ask how  BLG  fits in? To cut a long  story short \cite{Lambert:2010ji,Bashkirov:2011pt,Agmon:2017lga}:
\bigskip  
\begin{enumerate}[i)]
\item BLG $(SU(2)\times SU(2))/{\mathbbm Z}_2$ at $k=1$ is dual to ABJM $U(2)\times U(2)$ at $k=1$, i.e. 2 M2's in ${\mathbbm R}^8$
\item BLG $SU(2)\times SU(2)$ at $k=2$ is dual to  ABJM $U(2)\times U(2)$ at $k=2$, i.e. 2 M2's in ${\mathbbm R}^8/{\mathbbm Z}_2$
\item BLG  $(SU(2)\times SU(2))/{\mathbbm Z}_2$ at $k=4$ is dual to ABJ $U(2)\times U(3)$ at $k=2$, i.e. 2 M2's in ${\mathbbm R}^8/{\mathbbm Z}_2$ with torsion
i.e. 2 M2's in ${\mathbbm R}^8/{\mathbbm Z}_2$
\item BLG  $(SU(2)\times SU(2))/{\mathbbm Z}_2$ at $k=3$ is dual to ABJM $U(3)\times U(3)$ at $k=1$ without the centre of mass multiplet, i.e. the interacting part of 3 M2's in ${\mathbbm R}^8$ 
\end{enumerate}
\bigskip  
So BLG describes 2 or even 3 M2-branes in ${\mathbbm R}^8$ or ${\mathbbm R}^8/{\mathbbm Z}_2$ with all symmetries manifest (although not translations in the former case).

 \subsection{Lessons and questions}

So let me close the discussion of M2-branes with some lessons and a question. 
It is in general too much to ask for all symmetries to be manifest in the classical Lagrangian. In particular the true symmetries and dynamics only arise in the quantum theory using `quantum' operators, i.e. operators that are not constructed directly out of the fields and which do not have a classical analogue. Furthermore the role of the gauge group is very non-trivial and global issues matter. 

Lastly my question is: is there a role for the general BLG theories (i.e. for $k>4$)? They exist as maximally supersymmetric field theories which have a  weakly coupled limit as $k\to\infty$. Due to their moduli space they seem rather non-geometric but perhaps slightly deeper in the sense that one can find   the M2-brane theories by taking a ${\mathbbm Z}_k$ quotient of  them \cite{Lambert:2010ji}.

\section{M5-branes}

 The decoupling limit of $N$ M5-branes leads to an interacting CFT in 5+1 dimensions with an $SO(5)$ R-symmetry coming from rotations in the transverse 5-plane in eleven dimensions. In the Abelian case $N=1$ the dynamics are known \cite{Perry:1996mk,Howe:1997fb,Pasti:1997gx}.
 
The field content consists of five scalars $X^I$ (so now $I=6,7,8,9,10$ and $\mu=0,1,2,3,4,5$), a 2-form $B$ with self-dual field strength $H$ and a 16-component fermion  $\Psi$.
%\begin{itemize}
%\item five scalars $X^I$
%\item  two-form $B$ with self-dual field strength $H$
%\item 16-component fermion  $\Psi$
%\end{itemize}
At the linearised level we simply have
\begin{equation}
\begin{aligned}
\partial_\mu \partial^\mu X^I &=  0~,\\
H_{\mu\nu\lambda}   &=   3\partial_{[\mu}B_{\nu\lambda]}~, \qquad  H_{\mu\nu\lambda}=\frac{1}{3!}\varepsilon_{\mu\nu\lambda\rho\sigma\tau} H^{\rho\sigma\tau}~,\\
i\Gamma^\mu\partial_\mu\Psi &= 0\ .
\end{aligned}
\end{equation}
For $N>1 $  one finds the  interacting $A_{N-1}$ $(2,0)$ theory. 

\begin{figure}
\includegraphics*[width=\columnwidth]{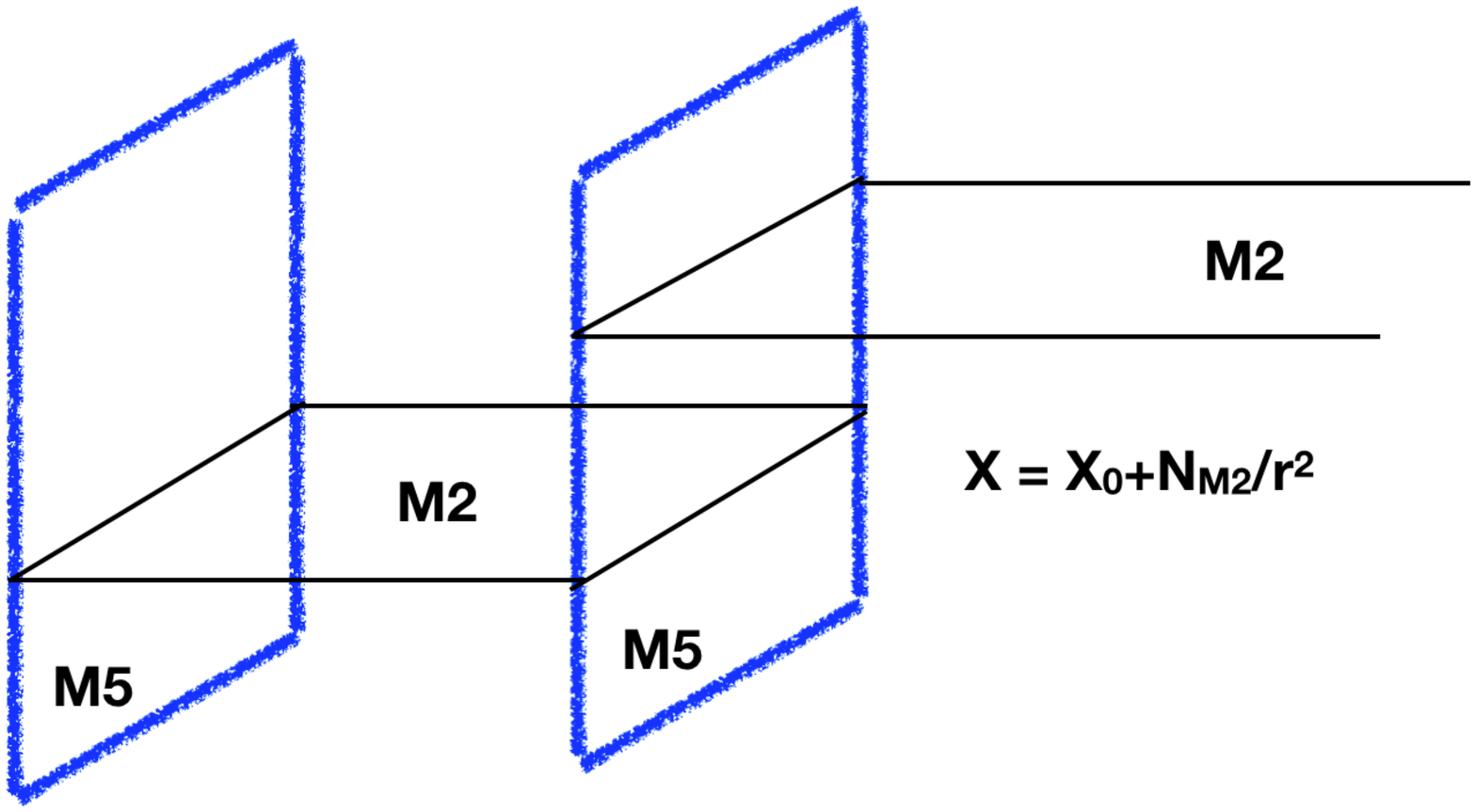}
\caption{\label{SDS}{M2 ending on M5's}}
\end{figure} 
 The dynamics are thought to arise from self-dual strings associated to M2-branes ending on M5-branes, just as D-brane dynamics arise from the end point of open stings (see figure \ref{SDS}). These are  the natural BPS states and 
Wilson-lines are replaced by surface operators. The 
 Abelian case has been long understood \cite{Howe:1997ue} however the 
 non-Abelian case of great interest as a higher gauge theory analogue of the Nahm transform \cite{Palmer:2013haa}.
 Finally we mention that 
 AdS/CFT predicts that the number of `degrees of freedom' of $N$ M5-branes scales as $N^3$ \cite{Klebanov:1996un}.

 \subsection{Reduction on $S^1$}
 
Let us wrap $N$ M5-branes on  $S^1$ of radius $R_5$. According to the M-theory  dictionary this leads to $N$ D4-branes in type IIA string theory with coupling $g_s = R_5/l_s$. These are in turn described by $U(N)$ 5D MSYM   and coupling $g^2 = 4\pi^2 R_5$.
 So the  $(2,0)$ theory is a UV completion of 5D MSYM with enhanced Lorentz symmetry \cite{Seiberg:1997ax}
\begin{equation}
SO(1,4)_L\times SO(5)_R \longrightarrow SO(1,5)_L\times SO(5)_R\ .
\end{equation}
This is another `prediction' of M-theory for quantum field theory: there exists a 6D SCFT that provides a UV completion of 5D MSYM.

In this story the Kaluza-Klein momenta  are carried by instanton-solitons $F=\pm\star F$ \cite{Rozali:1997cb}:
 \begin{equation}
 P_5 = \frac{n}{R_5}\ ,\qquad n = \frac{1}{8\pi^2}{\rm tr}\int F\wedge F\ .
\end{equation}
These states carry charges of  the topological current
 \begin{equation}
 J^\mu= \frac{1}{32\pi^2}{\rm tr}\int \varepsilon^{\mu\nu\lambda\rho\sigma}F_{\nu\lambda} F_{\rho\sigma}\ ,
\end{equation}
for which all perturbative states are uncharged.

\subsection{Reduction on ${\mathbbm T}^2$}

Let us reduce again on another $S^1$ with radius $R_4$. Here we find 4D $U(N)$ MSYM with coupling $g^2 = 2\pi R_5/R_4$. This theory has an S-duality that swaps perturbative modes with  monopoles and $R_4\leftrightarrow R_5$.  However from the 6D point of view 
this is a modular transformation of ${\mathbbm T}^2=S^1\times S^1$ which is a diffeomorphism   and hence is, or should be,  a manifest symmetry of the $(2,0)$ theory.\footnote{4D MSYM is only self-dual for ADE gauge groups so the  $(2,0)$ theory can only exist for ADE gauge groups. Indeed it was first constructed by a decoupling limit of type IIB on  $K3$  with an ADE singularity \cite{Witten:1995zh}. }

 \subsection {No Action?!}
 
 There are several arguments/challenges/opportunities\footnote{Delete as appropriate.} against constructing a 6D action for the $(2,0)$ theory. Let us discuss some.
 
 \begin{enumerate}[i)]
\item Even without worrying about self-duality there are no `good' interacting  Lagrangians in 6D. In particular the Lagrangian must take the form
\begin{equation}
\begin{aligned}
{\cal L}_{6D} \sim    &H_{\mu\nu\lambda}H^{\mu\nu\lambda} + D_\mu X^I D^\mu X^I\,+\\
 &+\, \underbrace{(X)F_{\mu\nu}F^{\mu\nu} + (X)^3}_{\rm unbounded} + \mbox{non-renormalizable}~,
\end{aligned}
\end{equation}
which is problematic as the  interactions are either non-renormalizable or unbounded or both.
 So what would the Lagrangian look like? In what space of classical theories does it exist?

\item How can one obtain   the 4D MSYM action which takes the form
 \begin{equation}S_{4D {\rm MSYM}} = \frac{R_4}{2\pi {  R_5}}\int {\rm d}^4x{\cal L}_{4D {\rm MSYM}}\ ,
  \end{equation}
 from the standard Kaluza-Klein result
 \begin{equation}S_{6D} =   
 \int {\rm d}^6x{\cal L}_{6D} = 4\pi^2 R_4{  R_5} \int {\rm d}^4x{\cal L}_{6D\, \mbox{\tiny zero-modes}}\ ,
 \end{equation}
since the dependence on $R_5$ is inverted between the two \cite{Witten:2009at}?

\item Let us consider dimensional reduction to ${\mathbbm R}^{1,1}$ on some ${\cal M}_4$. This leads to a 2D theory with $b^+_2({\cal M}_4)$ chiral bosons and $b^-_2({\cal M}_4)$ anti-chiral bosons.
However it is known that there is no modular invariant partition function if $\sigma({\cal M}_4)= b^+_2({\cal M}_4)-b^-_2({\cal M}_4)\notin 8{\mathbbm Z}$. On the other hand there is Rohlin's theorem which states that for compact 4D spin-manifolds $\sigma({\cal M}_4)\in 16 {\mathbbm Z}$. Thus it almost seems as if things should go the other way: the existence of a $(2,0)$ action would imply a weaker version of Rohlin's theorem (weaker by a factor of 2). However  one knows that non-spin manifolds, such as ${\mathbbm CP}^2$ with $\sigma({\mathbbm CP}^2)=1$,  can arise in M-theory and hence the M5-brane on ${\mathbbm R}^{1,1}\times  {\mathbbm CP}^2$ should make sense. But it cannot have an action, so therefore no diffeomorphism invariant action in 6D \cite{Witten:1996hc}.

\item   We have seen that the $(2,0)$ theory exists for ADE gauge groups but it is also known that reduction on $S^1$ with a boundary condition that twists  by an outer-automorphism  gives 5D MSYM with $B,C$ gauge groups. Thus if one had an action it should be subjected to  the {Tachikawa Test} \cite{Tachikawa:2011ch}:  given a  $SU(2n)$ $(2,0)$ theory action with an ${\mathbbm  Z}_2$ twist along $S^1$, does it give  $SO(2n+1)$ 5D MSYM?\footnote{NB: $SO(2n+1)\nsubseteqq  SU(2n)$.}
\end{enumerate}

 \subsection{Constructions}
 
 Let us now review some constructions of the $(2,0)$ theory that have been proposed. 
 
 \subsubsection{DLCQ}
   
 Consider null-compactification: $x^\pm = x^0\pm x^5,x^i$, $i=1,2,3,4$
 \begin{equation}
 x^- \cong x^- + 2\pi R_-\qquad {\rm and \ fix }\qquad P_- = K/R_-\ .
\end{equation}
We should view this as the limit of an infinite boost $v=1-\epsilon^2\to 1$ of a space-like compactification $x^5\cong x^5 + 2\pi R_5$ resulting in
\begin{equation}
R_- = R_5/\epsilon\ .
\end{equation}
To keep $R_-$ finite one must shrink $R_5\to 0$ and hence the $(2,0)$ theory on $S^1$ is well described  5D MSYM with fixed $P_5 = K/R_5$.
In this limit $K$ is given by the instanton number
 \begin{equation}
K = \frac{1}{8\pi^2 }{\rm tr}\int F\wedge F\ .
\end{equation}
 and we are looking at the sector of 5D MSYM with instanton number $K$.
Thus the  dynamics are reduced to quantum mechanics on the moduli space of $U(N)$ instantons with instanton number $K$ \cite{Aharony:1997th}.\footnote{NB:   This relies heavily on the fact that  $g^2 \propto R_5$ so we find  weakly  coupled 5D MSYM, which is a unique feature of the $(2,0)$ theory compared with compactifications of Lagrangian theories.}
    
  \subsubsection{Deconstruction} 
 
\begin{figure*}
\includegraphics[width=\textwidth]{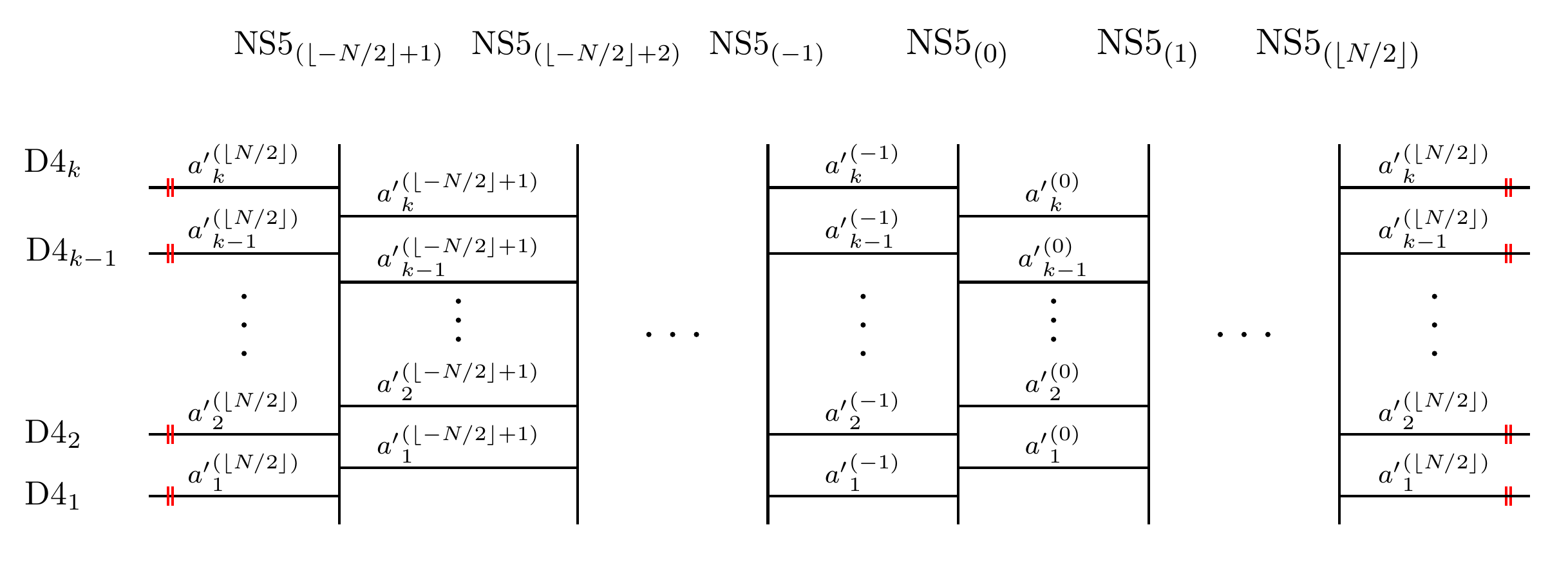}
\caption{\label{moose}{The (2,0) quiver}}
\end{figure*} 
   
Construct a quiver (moose) diagram  arising from the brane diagram in figure  \ref{moose}  (which was stolen from \cite{Hayling:2017cva}),
 where the left and right sides are identified into a periodic direction.
The D4-branes are described by $(SU(K))^N$ SYM with $N_f=2K$ fields in the bi-fundamental of each $SU(K)$.\footnote{NB: The roles of  $K$ here is the same as $N$ elsewhere in our discussion and the role of $N$ here has no analogue elsewhere.} This gives a 4D  ${\cal N}=2$ SCFT.

The next step is to go out on the Higgs' branch breaking $(SU(K))^N\to SU(K)$. A careful tuning of parameters: scalar vev's, coupling $g$ and number of nodes $N$ leads to a well-defined limit as $N\to\infty$.

The periodicity leads to a finite but large tower of states which for low energy look like a KK-tower.  However there is also  an S-duality of the quiver field theory so that the  KK-tower  is enhanced non-perturbatively to an $SL(2,{\mathbbm Z})$ multiplet of two towers. Thus as $N\to \infty$ one reconstructs a 6D theory with $SO(5)$ R-symmetry \cite{ArkaniHamed:2001ie}. 

This has recently been successfully used to make exact localization calculations \cite{Hayling:2017cva}.

\subsubsection{5D MSYM} 

Here the idea is that maybe 5D MSYM is actually well defined non-perturbatively, despite being perturbatively non-renormalizable,  and it is an exact description of the $(2,0)$ theory  on $S^1$ \cite{Douglas:2010iu,Lambert:2010iw}. In particular it contains a complete KK tower of soliton states so any UV completion would have to remove these and replace them with Fourier modes of some fields.  So why bother? Then one must hope that the
perturbative divergences are removed by small instanton-soliton effects \cite{Papageorgakis:2014dma}. In addition it seems that for this to work we need to include zero-sized instantons but one can see $N^3$ behaviour \cite{Kim:2011mv,Kallen:2012zn}.

In this scenario the extra momentum can be inserted by `instanton' operators \cite{Lambert:2014jna,Tachikawa:2015mha} 
\begin{equation}
\langle{\cal I}(y)_n{\cal O}(z)\cdots\rangle = \int_{\frac{1}{8\pi^2}{\rm tr}\oint_y F\wedge F = n}  D\Psi DA\ {\cal O}(z)e^{-S}
\end{equation}
which are analogous to the monopole operators that we saw before for M2-branes.

If so then 5D  MSYM does provide an `action' for the $(2,0)$ theory on $S^1$ for any radius. We note that if ${\cal M}_4=S^1\times {\cal M}_3$ then  $b^+_2({\cal M}_4)=b^-_2({\cal M}_4)$ and hence no-chiral modes, in agreement with the discussion above.

We could consider instead ${\cal M}_4$ as multi-Taub-NUT space with $b^+_2({\cal M}_4)\ne0$. This is non-compact but has a nontrivial $S^1$ fibration.   Reducing to IIA on the fibre leads to D4-branes intersecting with D6-branes. Here there are 2D chiral charged modes that are localised   where  fibration shrinks to zero size. The 5D MSYM that arises from reduction on a circle fibration has been discussed by  \cite{Linander:2011jy,Cordova:2013bea}. In this case one finds that the required chiral modes exist as solitons \cite{Ohlsson:2012yn,Lambert:2018mfb} and are described by a 5D version of a WZWN model. 
 
There is a related proposal where the $(2,0)$ theory on ${\mathbbm R}\times S^5$ is reduced to to 5D MSYM on ${\mathbbm R}\times {\mathbbm CP}^4$ with a Chern--Simons term \cite{Kim:2012tr}. In this case the coupling is related to the Chern--Simons level and so quantised. 

\subsubsection{Interrelations and other proposals}  

In fact these three descriptions are all related:
\bigskip  
 \begin{enumerate}[i)]
 \item
 The DLCQ description of the $(2,0)$ theory  must also give  the UV completion of 5D MSYM. But it only uses information arising from the classical IR dynamics of instanton-solitons of 5D MSYM. So somehow the IR behaviour of the theory is enough to determine its UV completion. This suggests that 5D MSYM is indeed well-defined without additional degrees of freedom. 
  \item 
The action obtained from deconstruction is a `lattice'-like  regularization of the 5D MSYM action \cite{Lambert:2012qy}. 
\item 
One cannot define 5D MSYM without also defining   the $(2,0)$ theory on $S^1$.
 \end{enumerate} 
 \bigskip
In addition to these proposals there also exist some action constructions in the literature. Let us list a few here:
\bigskip  
\begin{enumerate}[i)] 
 
\item  Twistor-inspired and higher gauge theory action \cite{Saemann:2011nb,Saemann:2017rjm}
 
\item  5D MSYM with KK-tower \cite{Bonetti:2012st}
 
\item  Mixed 5D/6D action \cite{Chu:2012um}
 
\item  $G\times G$ action \cite{Chu:2011fd}
 
\item  Non-local 6D action \cite{Ho:2011ni}

\item $(1,0)$ Lagrangians and tensor hierarchy \cite{Samtleben:2011fj}
 \end{enumerate}
 \bigskip  
 
 \subsection{Relations to M2-branes} 
 
   There are also a few ways that we expect M5s to arise from M2's.
   
\subsubsection{`T-duality'}
  
Here we consider a three-torus ${\mathbbm T}^3$ and reduce to IIA on the first $S^1$, T-dualize to type IIB on the second $S^1$, T-dualize back to IIA on the third $S^1$  and finally lift back up to M-theory, now on a dual $\hat {\mathbbm T}^3$. Using the standard rules we find that 
\bigskip
 \begin{enumerate}[i)]
 \item M5's on ${\mathbbm T}^3$  map to M2's orthogonal to $\hat {\mathbbm T}^3\times {\mathbbm R}^5$ 
 \item M2's orthogonal to ${\mathbbm T}^3$ map M5's on  $\hat {\mathbbm T}^3$
 \end{enumerate}
 \bigskip
 However we must take the decoupling (low energy) limit to isolate the M-brane field theories. This requires that we take $R\to 0$ in the original ${\mathbbm T}^3$ (so $\hat R\to\infty$ in the dual $\hat {\mathbbm T}^3$).

The first relation is rather trivial: the M5-brane on ${\mathbbm T}^3$ gives 3D MSYM and shrinking the torus takes it to strong coupling.  Thus we recover the M2-brane SCFT as a strong coupling IR limit of  3D MSYM.
 
The second relation says that  we can construct M5-branes by looking at M2-branes in a shrinking transverse ${\mathbbm T}^3$. However enacting this is more tricky because translational symmetry is not manifest in the M2-brane Lagrangian. An attempt was tried in \cite{Jeon:2012fn} and gives a modified version of 5D MSYM. 

\subsubsection{Flux background}

When M2-branes are placed in a background 3-form flux they expand into M5-branes on $S^3$ by the Myers effect. The resulting M5-brane action was constructed from the M2-brane action in \cite{Nastase:2009ny} but one  finds 5D MSYM on $S^2$. However when the monopole states in ABJM are included one finds that these map to instanton-soliton states 5D MSYM \cite{Lambert:2011eg}.

 \subsubsection{M2's with a Nambu bracket}

 There are infinite dimensional 3-algebras that can be used in the BLG theory. In particular the Nambu bracket 
 \begin{equation}
 [X^I,X^J,X^K] = \epsilon^{ijk}\partial_iX^I\partial_i X^J\partial_k X^K\ ,
 \end{equation}
is an example where $X^I$ are functions of some three-manifold $\Sigma$. It has been observed  that substituting this into BLG leads to an Abelian M5-brane wrapped on an auxiliary  $\Sigma$ \cite{Ho:2008nn,Bandos:2008fr}.

 \subsection{Open problems and wishes}

 Let me close this discussion of M5-branes with some open problems and wish list of results:
 \begin{enumerate}[i)]
 \item Provide a field definition/construction of the $(2,0)$ theory i.e. without recourse to String Theory or M theory
  \item Find the mathematical structures that best capture aspects of the $(2,0)$ theory e.g. Non-Abelian periods of 2-forms.  Twistors, Lie-2-Groups etc. 
   \item  Obtain  calculable formulations of the  $(2,0)$ theory with  6D Diffeomorphisms and Lorentz!
  \item Construct an action (?!), Partition function(s), families of actions or something action-like.
\item Better understand   `quantum operators' such mono\-pole and instanton operators.
\item Explore the relation between M2's and M5's more
 \item Make S-duality manifest?
 \item Make the $N^3$ behaviour more apparent
 \end{enumerate}

  \section{A representation of the $(2,0)$ superalgebra}

Finally in this last section I wanted to indulge myself by reporting on my own recent work that I hope is of interest to the conference crowd and I welcome any suggestions. In particular in 
\cite{Lambert:2010wm,Lambert:2016xbs} my collaborators and I constructed a representation of the $(2,0)$ superalgebra acting on a set of fields:
\begin{equation}
\begin{aligned}
\delta X^i &= i\bar\epsilon \Gamma^i\Psi~,\\
\delta Y^\mu & = \frac{i}{2} \bar\epsilon \Gamma_{\lambda\rho}{ C^{\mu\lambda\rho}}\Psi~,\\ 
\delta H_{\mu\nu\lambda} & = 3i\bar\epsilon \Gamma_{[\mu\nu}D_{\lambda]}\Psi + i \bar\epsilon \Gamma^i\Gamma_{\mu\nu\lambda\rho}[Y^\rho,X^i,\Psi]\,+\\ 
&\kern.5cm+\, \frac{i}{2}\bar\epsilon { (\star C)_{\mu\nu\lambda}}\Gamma^{ij}[X^i,X^j,\Psi]\,+\\ 
&\kern.5cm+\, \frac{3i}{4}\bar\epsilon \Gamma_{[\mu\nu|\rho\sigma}{ C^{\rho\sigma}{}_{\lambda]}}\Gamma^{ij}[X^i,X^j,\Psi]~,\\
\delta A_\mu(\cdot) & = i\bar\epsilon\Gamma_{\mu\nu}[Y^\nu,\Psi,\ \cdot\ ] +\frac{i }{3!} \bar\epsilon { C ^{\nu\lambda\rho}}\Gamma_{\mu\nu\lambda\rho}\Gamma^i[X^i,\Psi,\ \cdot\ ]~,\\  
\delta \Psi & = \Gamma^\mu\Gamma^i D_\mu X^i\epsilon + \frac{1}{2\cdot 3!}H_{\mu\nu\lambda}\Gamma^{\mu\nu\lambda}\epsilon\,-\\ 
&\kern.5cm-\,\frac{1}{2}\Gamma_\mu\Gamma^{ij}[Y^\mu,X^i,X^j]\epsilon\,+\\ 
&\kern.5cm+\,\frac{1}{3!\cdot 3!} { C _{\mu\nu\lambda}}\Gamma^{\mu\nu\lambda}\Gamma^{ijk}[X^i,X^j,X^k] \epsilon\ .
\end{aligned}
\end{equation} 
Here $X^I$, $\Psi$ and  $H_{\mu\nu\lambda}$  are dynamical taking values in a totally anti-symmetric 3-algebra,  $A_\mu$ and $Y^\mu$ are auxiliary   and 
{$C_{\mu\nu\lambda}$} is a  background (Abelian) three-form. Lastly we have
\begin{equation}
\Gamma_{012345}\epsilon=\epsilon~,\qquad \Gamma_{012345}\Psi=-\Psi\ .
\end{equation}

A standard (but trust me tedious) calculation shows that  
this system indeed closes on the following equations of motion (in this section we will omit fermions as much as possible for the sake of clarity)
\begin{equation}
\begin{aligned}
0&=\Gamma^\rho D_\rho\Psi + \Gamma_\rho\Gamma^i [Y^\rho,X^i,\Psi]\,+\\
&\kern.5cm+\,\frac{i }{12} C^{\rho\sigma\tau}\Gamma_{\rho\sigma\tau}\Gamma^{ij}[X^i,X^j,\Psi]~,\\
0&=D^2 X^i+ [Y^\mu,X^j,[Y_\mu,X^j,X^i]]\,+\\
&\kern.5cm+\,   \frac{1}{2\cdot 3!} C^2[X^j,X^k,[X^j,X^k,X^i]] + \mbox{fermions}~,\\
0&=D_{[\lambda}H_{\mu\nu\rho]}+ \frac12(\star C)_{[\mu\nu\lambda}[X^i,X^j,[Y_{\rho]},X^i,X^j]]\,+\\
&\kern.5cm+\,\frac{1}{4}\varepsilon_{\mu\nu\lambda\rho\sigma\tau}[Y^\sigma,X^i,D^\tau X^i] + \mbox{fermions} 
\end{aligned}
\end{equation}
as well as constraints:
\begin{equation}
\begin{aligned}
F_{\mu\nu}(\cdot)&=[Y^\lambda,H_{\mu\nu\lambda},\ \cdot\ ] -(\star C)_{\mu\nu\lambda}[X^i,D^\lambda X^i, \ \cdot\ ] +\mbox{fermions}~\\
0&=D_\mu Y^\nu -\frac{1}{2}  H_{\mu\lambda\rho}C^{\nu\lambda\rho}~,\\
0 & = [Y^\mu,D_\mu(\cdot),\ \cdot'\ ]  +\frac{ 1}{3}[D_\mu Y^\mu,\ \cdot\ ,\ \cdot'\ ]~,\\
0&= C^{\mu\nu\lambda}D_\lambda (\cdot) - [Y^\mu,Y^\nu,\ \cdot\ ]~,\\
0 &=C\wedge Y~,\\
0& = C_{ \sigma[\mu\nu }C^{\sigma}{}_{\lambda ] \rho}\ .
  \end{aligned}
  \end{equation}
  There is a conserved energy-momentum tensor:
  \begin{equation}
\begin{aligned}
&T_{\mu\nu} =\\ 
&\kern.1cm=\frac{\pi}{2}\langle H_{\mu\lambda\rho},H_{\nu}{}^{\lambda\rho} \rangle\,+\\    
&\kern.5cm +\, 2\pi  \langle   D_\mu X^i,D_\nu X^i\rangle- \pi  \eta_{\mu\nu}\langle D_\lambda X^i,D^\lambda X^i\rangle\,-\\
& \kern.5cm  -\, \frac{\pi}{2}\eta_{\mu\nu}\langle [Y_\lambda, X^i,X^j],[Y^\lambda, X^i,X^j]\rangle\,+\\
&\kern.5cm +\, \frac{2\pi}{  3!}\Big(C_{\mu\lambda\rho}C_{\nu}{}^{\lambda\rho}-\frac16\eta_{\mu\nu}C^2\Big)\,\times\\
&\kern1cm\times\, \langle [  X^i,X^j,X^k],[  X^i,X^j,X^k]\rangle\,+\\
 &\kern.5cm+\,\frac{\pi}{3!}C_{\mu\lambda\rho}(\star C)_{\nu}{}^{\lambda\rho}\langle[X^i,X^j,X^k],[X^i,X^j,X^k]\rangle\,+\\ 
 &\kern.5cm+\,\mbox{fermions} \ .
 \end{aligned}
 \end{equation}
One can also compute the supercurrent, superalgebra and central charges but lets not list those here.

Even I think this is an unconventional system and cannot decide if it is ugly (probably) or beautiful (possibly). But let us explore it.  

\subsection{M5-branes}  
 
Let us start with the case $C_{\mu\nu\lambda}=0$. Here $D_\mu Y^\nu=0$ and we can fix 
\begin{equation}
Y^\mu = V^\mu T^4\ .
\end{equation}
where $T^4$ is some generator of the 3-algebra   and $V^\mu$ a constant vector.  All components of the fields along $T^4$ become free - the 6D centre of mass (2,0) multiplet.  The remaining modes are acted on by an $\mathfrak{su}(2)$ gauge algebra. The constraint $[Y^\mu,D_\mu,\ \cdot\ ]=0$ implies that these modes only depend on the coordinates orthogonal to $V^\mu$. We also note that we can extend to any gauge group by taking a Lorentzian 3-algebra.
 
But there are still some choices for $V^\mu$.  

\subsubsection{Space-like $Y^\mu$}

First we take $V^\mu =2\pi R_5\delta^\mu_5 $.
The constraints then say that the remaining dynamical fields only depend on $x^0,\ldots,x^4$  and
\begin{equation}
F_{\mu\nu} =2\pi R_5 H_{\mu\nu 5}\ .
\end{equation}
The dynamical equations then all arise from the action  
\begin{equation}
\begin{aligned}
&S = -\frac{4\pi^2}{R_5}{\rm tr}\int {\rm d}^5x \Big\{\frac14 F_{\mu\nu}F^{\mu\nu} +\frac12D_\mu X^iD^\mu X^i\,-\\
&\kern3cm-\, \frac 14[X^i,X^j]^2\Big\}+\mbox{fermions}  \ ,
\end{aligned}
\end{equation}
i.e. 5D MSYM, corresponding to M5-brane on $S^1$ and KK-modes are instanton-solitons: 
\begin{equation}
P_5 = \frac{n}{R_5}~, \qquad n = \frac{1}{8\pi^2} {\rm tr}\int_{{\mathbbm R}^4} F\wedge F\ .
\end{equation} 

\subsubsection{Time-like $Y^\mu$}

Secondly we can set $V^\mu =2\pi R_0\delta^\mu_0$ (i.e. time-like). Now  $F_{\mu\nu} = 2\pi R _0H_{\mu\nu 0}$ and the
dynamical equations   all arise from  a 5D Euclidean MSYM. It is similar in form to the familiar MSYM but with some different signs but still with an $SO(5)$ R-symmetry, so we will not bother to write the action here.   Such a Euclidean theory with compact $SO(5)$ R-symmetry was noted by  \cite{Hull:1999mt} as a time-like reduction of the M5-brane. This is somewhat novel as typically Euclidean MSYM theories have non-compact R-symmetry. This one arises from reduction of super-Yang--Mills in $5+5$ dimensions. Just as the usual 5D MSYM secretly has an extra hidden compact dimension this field theory has  an emergent compact time \cite{Hull:2014cxa}.

\subsubsection{Light-like $Y^\mu$}

Note that we can also choose to set $Y^\mu =2\pi R_-\delta^\mu_-$ so $D_-=0$. Here we find 
\begin{equation}
F_{ij} = 2\pi R_- H_{ij-}\ ,
\end{equation}
and  self-duality of $H$ leads to self-duality of $F_{ij}$. Similarly $G_{ij} = 2\pi R_-H_{ij+}$ is anti-self-dual (but does not satisfy Bianchi).
The fields now depend on $x^+, x^i$, $i=1,2,3,4$. 

The dynamics can all be derived from the action  \cite{Lambert:2018lgt}
\begin{equation}
\begin{aligned} 
S & =  \frac{4\pi^2}{R_-}{\rm tr}\int {\rm d}^4x {\rm d}x^+\Big\{\frac{1}{2} F_{+i}F_{+i} - \frac{1}{2} D_iX^I D_iX^I\,-\\
&\kern3.5cm-\,\frac{1}{2}  F_{ij} G_{ij}\Big\}+\mbox{fermions}\ .
\end{aligned} 
\end{equation}

This is a novel field theory in 4+1 dimensions invariant under  16 supersymmetries, translations in space and time, $SO(4)$ spatial rotations and an   $SO(5)$ R-symmetry, but no boost symmetry.

Observe that $G_{ij} = 2\pi R_-H_{ij-}$ acts as a Lagrange multiplier imposing
\begin{equation}
F_{ij} = \star F_{ij}\ .
\end{equation}
This restricts the dynamics to motion on the moduli space of self-dual gauge fields.

The action reduces to a sigma model on the ADHM moduli space of fixed instanton number $n$ \cite{Lambert:2011gb}:
\begin{equation}
\begin{aligned} 
S & =  \frac12  \int {\rm d}x^+ \big\{g_{MN}(\partial_+\xi^M -L^M)(\partial_+ \xi^N- L^N )\,-\\
&\kern2cm-\, g_{MN}K^MK^N\big\} +\mbox{fermions}\ .
\end{aligned}
\end{equation} 
Here
$L^M,K^M$ are vectors on moduli space determined by the vev's of $A_+$ and $X^I$.

We can view a null choice of $Y^\mu$ as a limit of an infinite boost of a space-like $Y^\mu$ where we saw that the spatial momentum was $n/R_5$. Thus we are looking at an M5-brane with $P_- = n/R_-$.
This reproduces the DLCQ description of the dynamics of M5-brane.

\subsection{M2-branes}

Let us take now  turn on the constant three-form $C_{\mu\nu\lambda}$.

\subsubsection{Space-like $C_{\mu\nu\lambda}$}

First take $C_{345} = l^3$ non-vanishing.
The constraint 
\begin{equation}
[Y^\mu,D_\mu\ \cdot\ ,\ \cdot'\ ]  +\frac{ 1}{3}[D_\mu Y^\mu,\ \cdot\ ,\ \cdot'\ ]=0
\end{equation}
suggests setting $\partial_a=0$, $a=3,4,5$ and $Y^\alpha=0$, $\alpha=0,1,2$. 
In this case the constraint
\begin{equation}
C^{\mu\nu\lambda}D_\lambda (\cdot) - [Y^\mu,Y^\nu,\ \cdot\ ]=0
\end{equation}
implies 
\begin{equation}
A_a(\cdot)   =  -\frac{1}{2l^{3}}\varepsilon_{abc}[Y^b,Y^c,\ \cdot\ ] \ .
\end{equation}
From this the remaining constraints can be solved leading to
\begin{subequations}
\begin{equation}
\begin{aligned}
H_{abc} &= -\frac{1}{l^6}[Y_a,Y_b,Y_c]~,\\
H_{\alpha bc} & = -\frac{1}{l^{3}}\varepsilon_{bcd}D_\alpha Y^d~,\\
H_{\alpha \beta c}&= -\frac{1}{l^{3}}\varepsilon_{\alpha\beta \gamma}D^{\gamma}Y_c~,\\
H_{\alpha\beta\gamma} & =  -\frac{1}{3!l^6}\varepsilon_{\alpha\beta\gamma}\varepsilon^{abc}[Y_a,Y_b,Y_c]\ ,
\end{aligned}
\end{equation}
and 
\begin{equation}
\begin{aligned}
F_{\alpha a}(\cdot) &=  \frac{1}{l^{3}}\varepsilon_{abc}[Y^b,D_{\alpha}Y^c,\ \cdot\ ]~,\\
F_{ab}(\cdot) &= \frac{1}{l^{6}} [Y^c,[Y_a,Y_b,Y_c], \ \cdot\ ] \ .  
\end{aligned}
\end{equation}
\end{subequations}
 Let us write $X^a = l^{-3/2}Y^a$, then everything is derived from the action (taking $I=3,4,5,\ldots,10$)
 \begin{equation}
\begin{aligned}
S   &= \int {\rm d}^3x\Big\{  \langle  D_\alpha X^I,  D^\alpha X^I\rangle\,-\\
&\kern1.5cm-\,\frac{1}{6} \langle [X^I,X^J,X^K],[X^I,X^J,X^K]\rangle\,+ \\
&\kern1.5cm+\,\varepsilon^{\alpha\beta\gamma}(A_\alpha,\partial_\beta A_\gamma )-\frac{1}{3}\varepsilon^{\alpha\beta\gamma}(A_\alpha,[A_\beta,A_\gamma])\Big\}+\\ 
&\kern1.5cm+\mbox{fermions}\ .
\end{aligned}
\end{equation}
i.e.   BLG .

This is consistent with a T-duality along the directions of $C_{\mu\nu\lambda}$:
\begin{equation}
\begin{array}{ccccccc}
M5:&0&1&2&3&4&5
\end{array}
\overset{T_{345}}{\iff} \begin{array}{ccccccc}
M2:&0&1&2&&&
\end{array}
\end{equation}
 
\subsubsection{Time-like $C_{\mu\nu\lambda}$}
 
We can also take a `time-like' $C_{045} = l^3$.  This  leads to a   Euclidean M2-brane theory with $SO(2,6)$ R-symmetry. The Lagrangian is similar in structure to the normal maximally supersymmetric M2-brane case but with some funny signs so we won't bother to give it here. 

This is consistent with  \cite{Hull:1998ym} where a time-like T-duality of M-theory leads to M*-theory with signature $(2,9)$ 
\begin{equation}
\begin{array}{ccccccc}
M5:&0&1&2&3&4&5
\end{array}
\overset{T_{034}}{\iff} \begin{array}{ccccccc}
E3:& &1&2&&&5
\end{array}  
\end{equation}
And thus an E3-brane in this theory would indeed have $SO(2,6)$ R-symmetry.  
 
\subsubsection{Light-like  $C_{\mu\nu\lambda}$}
 
We can also take a null $C_{04+} = l^3$ \cite{Kucharski:2017jwv} which leads to a rather odd system. In particular 
the fields depend only on $x^+,x^1,x^2$ and now $Y^3,Y^4,Y^-$ are non-zero. Furthermore  $Y^-$ joins up with $X^i$ to form an $SO(6)$ multiplet which we denote by $X^I$. As before 
 $H_{\mu\nu\lambda}$ is largely determined in terms of $Y^3,Y^4,Y^-$ but
 self-duality implies $Z=Y^4+iY^3$ is holomorphic $\bar DZ=0$ where $z=x^1+ix^2$. Lastly 
 $H=H_{+z3} = iH_{+z4}$ is undetermined.
 
One finds that the dynamics can be obtained from the action  \cite{Lambert:2018lgt}
\begin{equation}
  \begin{aligned}
S &= \int {\rm d}^2x\, {\rm d}x^+\Big\{\frac{1}{4 }\langle D_+Z,D_+\bar Z\rangle + \langle D\bar Z,  \bar H \rangle +   \langle \bar DZ,H\rangle\,-\\
&\kern1cm-\,\langle DX^I,\bar DX^I\rangle  -\frac{i}{4}\langle D_+X^I,[Z,\bar Z, X^I] \rangle\,- \\  
&\kern1cm-\,\frac{1}{8}\langle [  X^I,X^J,Z] ,[  X^I,X^J,{\bar Z}] \rangle\,+ \\ 
&\kern1cm+\,\frac{i}{2}(A_+, F_{z\bar z}) + \frac{i}{2}(A_z, F_{\bar z +}) +\frac{i}{2}(A_{\bar z}, F_{+z})\,+\\  
&\kern1cm+\,\frac{i}{2}(A_+,[A_z,A_{\bar z}])\Big\}\\ 
&\kern1cm+\mbox{fermions} \ ,  
\end{aligned}
\end{equation}
where $\Psi_\pm = \frac 12(1\pm \Gamma_{05})\Psi$.
 
This is a novel field theory in 2+1 dimensions invariant under   16 supersymmetries, translations in space and time, spatial $SO(2)$ rotations
and an $SO(6)$ R-symmetry, but again no boost symmetry.

Note that $H = H_{+z3}$ acts as a Lagrange multiplier imposing
\begin{equation}
\bar D Z =0\ .
\end{equation}
Furthermore there is a Gauss Law constraint arising from the $A_+$ equation of motion:
\begin{equation}
 F_{z\bar z}(\ \cdot\ ) =   - \frac{1}{4 }   \left[ X^I, \left[Z, \bar{Z}, X^I  \right],  \ \cdot\ \right]+ \cdots\ .
\end{equation}
Thus the motion is constrained to the Hitchin moduli space.
 
As above we can view $C_{34-}$ as the limit of an  infinite boost along  $x^5$ of  the  $C_{345}$ case. Indeed the Hitchin-system gives rise to a momentum along $x^5$:
\begin{equation}
{\cal P}_5 \sim \oint \big\{   \langle Z,\bar D\bar Z \rangle {\rm d}z
+     \langle \bar Z, D Z \rangle {\rm d}{\bar z}\big\}\ ,
\end{equation}
which appears as a winding of the M2-branes around $x^3,x^4$.
\begin{figure*}
\hspace{2.5cm}\includegraphics[width=13cm]{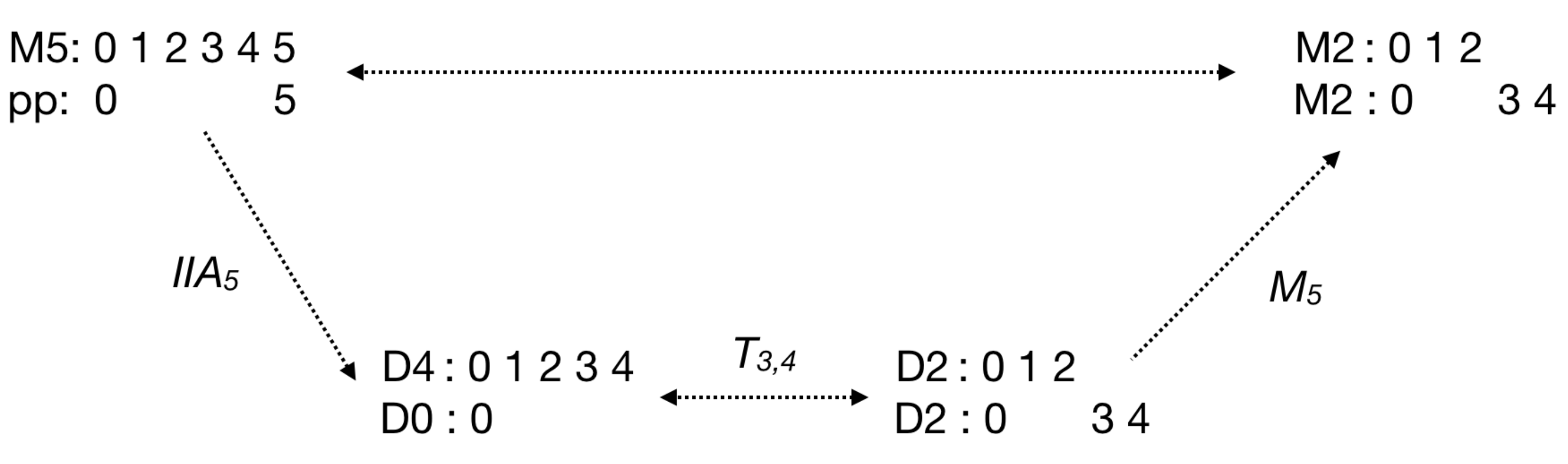} 
\caption{\label{MT}{T-duality in M theory}}
\end{figure*} 

So we are looking at intersecting M2-branes that have been boosted along $x^5$. This is a T-dual relation to momentum modes of the M5-brane (see figure \ref{MT}).
 
 \subsection{Observations and provocations} 
 
 So our representation of the $(2,0)$ superalgebra gives various field theories associated to M-branes.
\bigskip  
\begin{enumerate}[i)]
\item  5D MSYM as the M5 on $S^1$ 
\item  Maximally supersymmetric M2 branes
\item  Null M5-branes: QM on instanton moduli space
\item  Null intersecting M2-branes: QM on Hitchin moduli space
\end{enumerate}
\bigskip  
The later two are novel non-Lorentz invariant field theories whose on-shell dynamics reduces to one-dimensional motion on moduli space and breaks 1/2 the supersymmetry.

 The field theories that we obtain from this system are all consistent with the notion of `T-duality' (really a U-duality) in M-theory on ${\mathbbm T}^3$ along  $x^\mu, x^\nu, x^\lambda$ with radii $R_\mu, R_\nu, R_\lambda$ and 
\begin{equation}
C_{\mu\nu\lambda} = (2\pi)^3  R_\mu R_\nu R_\lambda\ ,
\end{equation}
but one needs to generalise all this to more than two branes in order to make it more concrete!

 This $(2,0)$ system is reminiscent of doubled field theory where $X^I$ is a  position coordinate and  $Y^\mu$ is a winding coordinate. Under T-duality along $x^\mu$ the corresponding 
$Y^\mu$'s become position coordinates. Furthermore 
 the $Y^\mu D_\mu =0$ constraint is like a section condition.
 Although it should be noted that the fields are only functions of ordinary 6D coordinates $x^\mu$ (i.e. not of the winding coordinates). It would be interesting to see if there is a deeper geometrical significance to the various constraints of the $(2,0)$ system.

\makeatletter

\interlinepenalty=10000  

\bibliography{allbibtex}

\bibliographystyle{prop2015}

\makeatother

\end{document}